\documentclass[12pt]{elsart}
\usepackage{lineno}%
{\par\runninglinenumbers
\usepackage{epsfig}
\abovedisplayskip=6pt
\belowdisplayskip=6pt
\textwidth=150mm
\textheight=220mm
\oddsidemargin=0.75 cm
\evensidemargin=0.75 cm

\begin{document}
\baselineskip 20pt plus .1pt  minus .1pt \pagestyle{plain}
\voffset +1.0cm \hoffset 0.0cm \setcounter{page}{01} \rightline
{} \vskip 1.0cm

\begin{center}
\Large {Clouds, solar irradiance and mean surface temperature over the last century}
\end{center}
\vspace {0.5cm}  
\begin{center}
\footnote{Corresponding author at P.N.Lebedev Physical Institute, Moscow, Russia\\
E-mail address: erlykin@sci.lebedev.ru (A.D.Erlykin)} 
A.D.Erlykin$^{(1,2)}$, T.Sloan$^{(3)}$, A.W.Wolfendale$^{(2)}$
\end{center}

\begin{flushleft}
(1) {\em P. N. Lebedev Physical Institute, Moscow, Russia}\\
(2) {\em Dept. of Physics, University of Durham, Durham, UK}\\
(3) {\em Dept. of Physics, Lancaster University, Lancaster, UK} 
\end{flushleft}

\begin{abstract}
The inter-relation of clouds, solar irradiance and surface temperature is complex and subject to different interpretations.  Here, we continue our recent work, which related mainly to the period from $1960$ to the present, back to $1900$ with further, but less detailed, analysis of the last thousand years. The last $20$ years is examined especially.  Attention is given to the mean surface temperature, solar irradiance correlation, which appears to be present (with decadal smoothing) with a $22$-year period; it is stronger than the 11-year cycle correlation with one year resolution.   UV in the solar radiation is a likely cause. Cloud data are taken from synoptic observations back to 1952 and, again, there appears to be a correlation - with opposite phase for high and low clouds - at the $20$-$30$ y level.
Particular attention is devoted to answering the question, `what fraction of the observed increase in mean Global temperature ($\sim 0.7^{\circ}$C) can be attributed to solar, as distinct from man-made, effects?'  We conclude that a best estimate is 'essentially' all from $1900$ to $1956$ and less than 14\% from $1956$ to the present.
\end{abstract}
\section{Introduction}
The undoubted `global warming' over the past half century or so has focused attention on the role of changes in solar irradiance (and the solar wind) on a variety of timescales and the relevance of cloud cover (CC).  As is well known, the effect of `solar forcing' on the Earth's climate is not fully understood (eg Foukal et al, $2004$; $2006$).  In particular, the  observed temperature changes are greater than would have been expected, so this is one reason for yet another examination of the problem.  Another is our analysis of the maps of Voiculescu et al ($2006$) in which we were unable to find a good meteorological reason for the observed geographical pattern of the regions having strong cloud cover (CC), solar irradiance (denoted SI) correlations.  As is well known, (eg Kristjansson and Kristiansen, $2000$ and Erlykin et al $2009$a), the observed CC, Sunspot Number (SSN) - which can be taken in first order as a proxy for SI - correlation is most unlikely to be due to cosmic ray variations, as proposed by a number of authors, and SI variations are favoured.  In what follows, we refer to `SI' but are mindful that the closely related solar wind may be the operative agent, instead.  The possible distinction is taken up later. Here, we examine the variations over the $20$th Century of both temperature and SI (via SSN) and CC.  Data on `temperature versus time' are available for most of the $20$th Century for many points on the Earth's surface (eg Hegerl et al, $2007$) and sunspot records are readily available over an even longer period. A complementary study is made, albeit with less rigour, of the last thousand years and, in view of the contemporary significance of the extended period of low (zero) sunspot numbers, the last $20$ years is examined in some detail.

The biggest problem relates to cloud cover.  Clouds are inevitably connected with temperature change and their relationship to surface temperature is cloud-height dependent.  Satellite data have been available since $1983$, only, and even here there are calibration uncertainties (eg Norris, 2000).   Cloud data over a longer period, post-$1952$, are, however, available from the synoptic reports summarised by Norris ($1998$, $2004$).  These relate to `upper' and `low' clouds and cover specific latitude ranges : $30^{\circ}$ S - $30^{\circ}$ N (ocean) and $30^{\circ}$ N - $60^{\circ}$N (ocean), and these data are used here.

The main thrust of the paper is to study the correlation of changes in solar irradiance with `climate' (temperature and clouds) for the various time periods from the standpoint of both the $11$-year and $22$-year cycles.  We aim to check that the temperature variations are in fact excessive and, in particular, to study the reason for the $22$-year cycle being so much stronger than that for the $11$-year cycle (as observed already by Miyahara et al, $2008$).

We then go on to determine a best estimate of the contribution of `natural' (SI) effects to the well known increase of Global temperature since $1900$. It is appreciated that others have also examined this topic but an independent study is clearly desirable.

\section{Temperature versus Solar Irradiance}
\subsection{An Overview}

Figure $1$ shows a collection of estimates of Global average surface temperature changes ($\Delta$T) versus estimated changes of SI ($\Delta$SI) from (mainly) previous work, in order to `set the scene'.  Details are given in Table $1$.  The data here are all from the analyses of others, except for $7$ and $3$.  Concerning `$7$' this is a straightforward plot of the variation of SI over the year for latitude bands and the corresponding summer/winter temperature difference.  Although a naive approach it has most of the ingredients for a realistic expectation for the rate of change of $\Delta$T with $\Delta$SI. It is extrapolated linearly downwards.  Another prediction is the line indicated `model'.  This was derived using the relation $\frac{\Delta  \textrm{T}}{\textrm{T}} = \frac{1}{4} \frac{\Delta \textrm{SI}}{\textrm{SI}},$ which fits the data for almost all planets and is as expected from evaluation of the equilibrium temperature of an insulated black surface normal to the Sun.  It is appreciated that, understandably, the planets do not satisfy the condition of an `insulated black surface' but there is just a systematic off-set of this equilibrium temperature and the actual (bright-side) temperature.  The off-set for most planets is a reduction of actual below equilibrium temperature of about $30\%$.  The exception is Mercury, having no atmosphere.  In the case of the Earth the ratio is a little higher, the equilibrium temperature being about $394^{\circ}$K and the actual $295^{\circ}$K (ie a reduction of about $25\%$).

In Figure $1$, we use T = $294^{\circ}$K, the equilibrium temperature, so the line is somewhat of an upper limit.  The line is approximately $\Delta$T($^{\circ}$C) = $\frac{\Delta SI}{SI}$ (in \%) ie at our datum $\Delta$T = $0.1^{\circ}$C, $\frac{\Delta SI}{SI} = 0.1\%$. The line is clearly just a datum in that feedback effects (positive or negative) can cause differences. This topic is taken up again later.

The point marked $3$ is from our recent paper (Erlykin et al, $2009$a).  In this paper we pointed out that the change in temperature since $1956$ about the smooth trend follows closely the change in SI (both averaged over the $11$ year solar cycles).  The change in the cosmic ray rate was also observed to follow the trend but is delayed by $\sim 2-4$ years.  We assume that the observed change in temperature is caused by the change in SI to give point $3$ on Figure $1$.

Point $6$ differs from the others in that it is from calculation of the effect of a change of distance of the Earth from the sun; the others all relate to measured, or inferred, temperature changes ($\Delta$ T) and associated changes in sunspot number ($\Delta$SI/SI).

Inspection of the Figure shows that with the exception of `$7$', the Global seasonal variations, the $\Delta$T values are higher than would have been expected from `expectation', a result of importance in view of the need to know the magnitude of the solar forcing at the $\Delta$SI $\simeq$ 0.1$\%$ level as a help to understanding fully the cause of temperature changes in general and Global Warming in particular.

One of the objectives of the present work is to attempt to clarify the situation of these small irradiance changes and to confirm, or otherwise, the `excess' values of $\Delta$T. Another is to endeavour to identify the actual cause of the temperature changes : SI as such or another phenomenon connected with the solar wind and to go on to determine the `natural' contribution to the $0.7^{\circ}$C Global Warming since $1900$.  Another objective is to examine the role of clouds, particularly from the standpoint of the $22$-y cycle.

\subsection{The last half-century - our own earlier work}
As remarked, we have already examined temperature, SI and CR data since $1956$ (Erlykin et al ($2009$a)).  It was concluded that the CR time variations did not fit the temperature variations and this was yet another reason to disbelieve a significant role for CR in generating clouds and thereby affecting temperature (other work includes Sloan and Wolfendale, $2008$). The conclusions in the post-$1956$ data related largely to the information from the large dip in $\Delta$T in the region of $1970$, which coincided with the well known low sunspot maximum (for Cycle $20$) in that year in comparison with the neighbouring Cycles (Figure $2$).  Next, we go further back to the beginning of the Century and see whether there is confirmation of the conversion from SI change to temperature change derived there is confirmed (point $3$ in Figure $1$).

\subsection{The temperature record for the last century}
\subsubsection{The mean over the Globe}

A disconcerting feature is the fact that the profile of surface temperature versus time is not unique but varies from place to place, not only in absolute magnitude but also in shape.  Figure $3$a shows the average surface temperature over land (LAN), over ocean (OCE) and averaged for the whole earth (GLO).  (The results are from Hegerl et al, $2007$). Inspection of the profiles of surface temperature vs year, from $1910$ to $2000$ for the $22$ regions distributed over the Globe (each having an area of $\sim 10^{7}$ km$^{2}$) shows not only differences in the temperature rise from place to place, but other differences, too.  Most pronounced is the movement of the peak at $1940$ in Figure $3$a which, although stable between ocean and land, is variable from one region to another.  Specifically, there are $12$ regions with peaks between $1938$ and $1942$ and $8$ regions with peaks between $1948$ and $1952$.  The cause of the dichotomy is probably the phenomenon encountered in the Altai region where the identification by Eichler et al ($2009$), of a $10-30$ year lag between solar forcing and temperature response, led the authors to postulate an indirect sun-climate mechanism involving ocean-induced changes in atmospheric circulation.  Having said that, it must be remarked that the $3$ regions nearest to Altai (`NAS', `CAS' and `TIB') all had peaks in $1940$, rather than being delayed by 10 years to $1950$.  In any event, the stability of the profiles in Figure $3$a for land (LAN) and ocean (OCE) is plain to see.

The role of the oceans in comparison with that of the land can be seen by means of the overall increase in temperature over land (LAN) being significantly bigger than that over the oceans (OCE).  Having said that, it is not clear why the `structure' in OCE is greater than that in LAN; although the peak at $1940$ and the dips at $1950$ - $1970$ are close in time, they are sharper in OCE.  Presumably the answer lies in the fact that there is a greater homogeneity of `ocean' than `land', the latter having a wide variety of terrains : industrial areas, farm lands, lakes, deserts, etc, all with different albedos and other properties.

In this connection it is necessary to consider further the possible time lags that can occur between SI changes and subsequent changes in $\Delta$T.  For a start, differences might be expected between ocean and land, in view of the different specific heats ($5$:$1$) and thermal conductivities.  However, inspection of Figure $3$a shows that the differences between LAN and OCE are only in amplitude and not in temporal position as mentioned already.  Thus, the strong $1940$s peaks are in the same year, as are the minima at $\sim$ $1950$ and $1969$.  That the near-surface air temperature has a very short time lag is evident from a night/day comparison; it is invariably `cold at night'.  Our earlier work (Erlykin et al, $2009$a)for the $11$-year averages showed a lag less than a year (indeed the best estimate would appear to be about minus one year, with respect to the SSN!).

In what follows for `mean Global temperature' we use the values from Hansen et al ($2006$), with an $11$-year smoothing to eliminate the first order solar-cycle variation, as adopted in our earlier work(Erlykin et al, $2009$a).

\subsubsection{The $1940$ peak}

It has been remarked already that there is a consistent peak in temperature in $1940$, for the Global averages, for all situations : GLO, LAN and OCE.

Inspection of Figure $2$, the sunspot number versus time for the last century, surprisingly shows no expectation of a maximum for $1940$, rather the envelope of the SS number leads to an expectation of a maximum $\Delta$T in about $1959$ if there is a causal connection between solar activity and temperature.  Here, there is, indeed, a small peak in OCE ($\simeq$ 0.04 $^{\circ}$ C) and GLO ($\simeq$ 0.03$^{\circ}$ C) but this is dwarfed by the $1940$ peak.

The surprising peak in $1940$ has been commented on by a number of workers (eg Hegerl et al, $2007$).  In the period $1930$ - $1960$, volcanic aerosols are thought to have had a negligible effect on Global temperatures (Lean et al, $2005$) but this is not the case with the ENSO tropical temperature index, which has a peak-to-peak temperature excursion of about $0.15^{\circ}$ C for $1$ y binning (and the well-known $2$ - $3$-y oscillation).  There was, in fact, a particularly strong El Nino in the period $1940-42$ but the geographical distribution of the `$1940$-peak', which not only varies by about 10 years, does not accord with expectation.  The detailed correction of the surface temperature for ENSO, volcanic aerosols, greenhouse gases and tropospheric aerosols by Lean et al ($2005$), although including a `peak' of about $0.05^{\circ}$, and a width of half height of $\sim$ $6$ years, still leaves an excess $\Delta$T of $\sim$ $0.2^{\circ}$C over nearly $10$ years unaccounted for.

Another explanation put forward is that the $1940$ `peak' was due to a post $1940$ `dip' caused by the effects of bio-mass burning (Nagashima et al, $2006$).

The detailed discussion of the `$1940$-peak' is to draw attention to the hazards involved in separating out a (small) solar forcing signal in the presence of other forcings, of inevitably uncertain magnitude.

\subsubsection{Temperature, SSN correlation over the whole Century}

In an attempt to apply a consistent analysis to the data we inspect the temperature record in Figure $3$b, (the upper line) which is decadely averaged and corrected for the long-term trend (as in Erlykin et al, $2009$a) and endeavour to correlate the patterns of $\Delta$T and SSN.  In view of the $11$-year smoothing of the SSN there are minima at the even Solar Cycle numbers indicated in the Figure.  It is evident that with the exception of the `$1940$-problem' there is a generally good correlation between the patterns and that $\Delta$T and SSN are causally correlated.  In order to derive a value to add to Figure 1 we examine the magnitudes of the dips in both $\Delta$T and SSN assuming that the two are strongly correlated.  Times are identified which are near the peaks of the (smoothed) SSN - identified by small vertical arrows - and the SSN dip determined for each of the $5$ regions (by `dip' we mean the mid point value with respect to the mean of the two end points).  The same end points were taken for $\Delta$T and the temperature dips found in a similar fashion.  The prominent dip marked by the $3$rd arrow in Figure $2$, ie Cycle $20$, is that studied in our earlier work (Erlykin et al, $2009$a).

We have taken the data from Figure 3b and measured off 3 points in each of the 5 dips, equally spaced.  In each case $\Delta$T and $\Delta$SSN are determined with respect to the chord joining the peaks in $\Delta$T corr.  Figure 4 shows the values so derived.  It will be observed that there is a very approximate linear dependence of $\Delta$T on $\Delta$SSN.  The slopes can be converted to $\Delta$T vs $\Delta$SI/SI using the conversion from Sloan and Wolfendale (2008) that $\Delta$SSN = 150 corresponds to the $\Delta$SI/SI = -0.07\% from Lean et al (2005).  The result is that $\Delta$T = (0.1 $\pm$ 0.03)$^{\circ}$C corresponds to $\Delta$SI/SI = (0.014 $\pm$ 0.004)\%. Eliminating points `30' and `50' in Figure 4, ie the data containing the dips in 1930 and 1950, which might be justifiable because of problems with the 1940 peak, can be seen to have no effect on the overall slope.  The point is plotted as `10' in Figure 1.

It should be remarked that SI (taken from the work of Wang et al (2005)) follows SSN rather closely from 1956 to 1992 but there is divergence thereafter; this is another indicator that Cycle 23 is anomalous.

As a check, we have used an alternative set of Global temperature data, that over land alone and for the two Hemispheres separately (Peixoto and Oort, 1992).  The averages to 1985 are: $<\Delta T>$ = $0.15 \pm 0.05^{\circ}$ C for the N hemisphere, and 
$<\Delta T>$ = $0.05 \pm 0.02^{\circ}$ C for the S hemisphere.

The Global mean (0.06 $\pm$ 0.02$^{\circ}$C) is consistent with our 0.05 $\pm$ 0.02$^{\circ}$C just derived.

The fact that the mean value of $\Delta$T over land is greater than that over the oceans is understandable in view of the higher thermal inertia of the oceans (see 2.3.1 and Figure 3a).  The value of $\Delta$T over land in the S Hemisphere is presumably lower than that in the North for reasons of the greater proximity to water in the South.

We are mindful of the contributions of other effects to changes in surface temperature : ENSO, volcanoes, ozone and greenhouse gases.  Concerning the last mentioned, the smoothed contribution is included in the slow systematic change in $\Delta$T with time; it is the shorter term variable part which is of concern here.  Estimates of the necessary corrections have been made by us using data from Lean et al (2005) and Mechl et al (2004).  The values were smoothed over 5 year intervals and have been applied to the upper line in Figure 3b to give the middle curve, labelled $\Delta$T$_{corr}$.  The median spread in corrections for the 5-yearly smoothing is $\pm 0.05^{\circ}$ C.  Our estimate for the 1-year smoothing is $\pm 0.09^{\circ}$C; this value will be needed later.

Repeating the analysis to give the mean values of $\Delta$T and SI for the `corrected' time dependence in Figure 3b gives $\Delta$T = 0.01$^{\circ}$C $\pm$ 0.02$^{\circ}$C, for $\frac{\Delta SI}{SI} = 0.013\% \pm 0.003\%$ ie very similar to the earlier value. It is reassuring that the perturbing factors (ENSO, volcanoes, etc) do not invalidate our analysis.

The conversion value for 1956-2001 derived by us in Erlykin et al (2009a), of $\frac{\Delta SI}{SI} = 0.015\%$ for $\Delta$ T = $0.1^{\circ}$C, is therefore confirmed within the uncertainties.

\subsubsection{Temperature, SSN correlations over the last 1,000 years}

Although there are no direct SSN measurements over the whole of the millennia, proxy indicators of the SI have been used.  For example, Crowley (2000) has used cosmogenic isotopes, specifically $^{10}$Be in ice cores, residual $^{14}$C from tree ring records and an estimate of $^{14}$C from $^{10}$Be fluctuations.  The same author derived a temperature record for the Northern Hemisphere, using instrumental data after 1860 and a proxy record prior to this date, the proxy being tree rings, corals and ice cores.  It is appreciated that there are many uncertainties for a time period of such length but we would contend that an analysis of correlations in the proxy data has some value.

Crowley gives the resulting `observed' temperature variation (we call it $\Delta$T (observed) and the expected, from the inferred $\Delta$SI temperature relation derived above, (we call it $\Delta$T (predicted) versus time from the year 1000 to 2000.  The data have been used by us to study the correlation for each century: 1000 - 1100.... to 1800-1900 with the results for the correlation parameter, p, and the slope of the line for $\Delta$T (observed) vs $\Delta$T (predicted) shown in Table 2.  It is interesting to note that the p-values are very small (ie the correlation is very significant) for the periods for which volcanoes contributed significantly to the $\Delta$T-value (1200-1300; 1400-1500 and 1800-1900), ie the corrections must have validity. Taking all the 1000 year data together and plotting $\Delta$T (obs) vs $\Delta$T (pred), we find a straight line fit with slope $0.77 \pm 0.05$, a correlation coefficient $r = 0.98$ and a correlation probability $p < 0.001$.

The overall situation regarding the probabilities is satisfactory and supports the contention that there is a good correlation over 1000 years.

The near-proportionality of $\Delta$T to the solar forcing ($\Delta$SI) used in the calculations, with the small volcanic forcing correction adds validity to the arguments put forward in section 2.3.3 for the last 50 years.  It remains to examine reason for the difference in the 11-year and 22-year temperature - SI relations.  That there are dependences with both 11-year and 22 year components is evident from many workers for the 1-year averaged direct solar cycle (eg Lean et al, 2005).

Concerning the 22-year (approximately) variation over the century, this follows directly from the present work (Figure 3b) for the last 100 years.  It is also present for the same period in (Erlykin et al, 2009b).  In that work we found for the Fourier frequency spectrum, peaks at 0.0039 and 0.0072 month$^{-1}$, i.e. 21- and 11.6 - years.

Independent analyses have been made by others (eg Vecchio and Nanni, 1994, for the last century) show peaks in the `relative variance' at $\sim$20 years (the Hale cycle).

Turning to the last millennium, inspection of the $\Delta$T, time profile shows the presence of some 50 peaks in this period, ie a mean separation of $\sim$20 years.  This is borne out by the Fourier analysis which shows a peak at about 20 years.

At the level of examination here there is no evidence for significant phase lag between the changes in SI and temperature, except, perhaps for the periods 1000 - 1100 and 1600 -1700, where the correlation probabilities in Table 2 are poor.

\subsection{Discussion of the temperature versus solar irradiance results}
\subsubsection{General Remarks}

Despite the fact that most of the `points' in Figure 1 are `high' - it does appear there is evidence in their favour.  Additional evidence comes from the fact that there is an upward progression of $\Delta$T with length of time over which the averaging is made, at least for those observations for which we have made the analysis, viz points: 7, 3, 10 and the point from the exhaustive study by Lean et al, (2005): 2. Of the others, 6 relates to a first order calculation of the expected effect of changes in the sun-earth distance and the others are approximate.

The `preferred' points are thus 7 (1-year), 2 (11-year), 3 and 10 (22-year).  There are now two questions: is the progression `reasonable'? and, what is the reason for the excess values of $\Delta$T over the Model expectation? Such a behaviour is not completely understandable.  Thermal inertia per se appears to be ruled out.  The datum line (denoted `7') relates to a yearly period and it seems that some 20-30 year is needed for this inertia to be largely overcome.  The inertia must be `resistive' in character (in part, at least); a `capacitive' component would give a phase-lag, which seems not to have been observed for the earth as a whole.  It is not self-evident that such a time is correct particularly because the `Model' line should be an equilibrium value and it is `low';presumably some form of positive feedback is operative.  The comments of Shindell et al (1999), with regard to the `disproportionate effects of UV' changes on the upper wind patterns are relevant, as in the model of Haigh (2007) and elsewhere, as will now be described.  Many workers have found increased values of $\Delta$T at heights well above ground level.  Specifically, Hood (1997), Hood and Soukharev (2000) and Gray et al (2009) have derived $\Delta$T values higher than the 0.1$^{\circ}$C ground level value, by a factor $\sim$4 at a pressure of 10hPa rising to $\sim$20x at a pressure of 0.1hPa.  These high values arise from the large 11-year cycle in solar UV and include positive feedback effects due to ozone.  At 0.1 hPa (65 km) the $\Delta$T change is seen to be some 50 times the Model prediction of Figure 1.

Referring to Figure 3b, it appears that the dips at 1910, 1930, 1970 and 1991 (ie the `22-year cycle'), are reflections of the fact that the peak SSN values alternate from Cycle to Cycle. A useful factor is the ratio of the sunspot numbers (monthly averages) for an even-numbered cycle to the mean of the adjacent ones. The ratios are: Cycle 14 : 0.82, Cycle 16 : 0.78, Cycle 18: 0.97, Cycle 20: 0.79 and Cycle 22 : 1.03. This last-mentioned arises because the peak at 2002 (cycle 23) is anomalously low; indeed cycle 23 is anomalous in many ways (see Section 5). This aspect will now be examined.

\subsubsection{`Solar irradiance'}

Although in first order, the change in Solar Irradiance, `SI', is proportional to sunspot number, `SSN', there are subtleties.  These arise from the spectral shape of the solar radiation.  It is well known that, although the fraction of the energy content of SI falls with increasing frequency, its 11-year cycle increases in amplitude.  Thus, there is the possibility that the pattern in Figure 3b is a consequence of UV as distinct from visible radiation.  Inspection of available data indicates typical radiance changes over the 11-year cycle of 0.15 Wm$^{-2}$ in the range 200 - 300 nm which, at the high altitudes at which this radiation is absorbed (above 20 km) is very large, considering the very low air density involved (the total, for all wavelengths, is only $\sim$1Wm$^{-2}$).

If the odd-even Cycle differences found for SSN (and other indicators) are present in the UV, too, then UV irradiance is a good candidate for the observations in Figure 3b.  This aspect can be considered further.

UV data are only available from 1978 (Viereck and Puga, 2005; Deland and Cebula, 2008), but can be extrapolated back just two years to the SSN minimum in 1976, yielding reasonable results for all 3 Cycles: 21, 22 and 23.  Integration over the UV intensity vs time for each Cycle gives (UV(22)/$<$UV(21,23)$>$~= 0.78, without doubt less than the (anomalous) ratio for SSN (1.03).

If the feature is common, viz that the UV (Even Cycle) is always significantly less than the mean of its neighbours, to a greater extent than is usually true for SSN, then we would have a ready explanation of the $\Delta$T value for the 22-year mean being proportionately greater than for the 11-year cycle (Figure 1, ie the black summary point from Figure 3b being `higher', with respect to either line than the 11-year points 1 and 2).

The role of UV in probably explaining the high value of $\Delta$T with respect to the Model prediction has already been referred to in 2.4.1.  It has relevance, too, to the 22-year cf 11-year difference.  In the work of Haigh (2007) the energy (excess) in the 11-year Cycle is deposited at about $50$ hPa, ie $20$ km altitude where 300 nm radiation is absorbed (this is one optical depth from the top of the atmosphere).  The air density here is $\sim 10^{-4}$ of that at ground level so that it is not surprising that the predicted temperature effect at this altitude is so great. The model involves a perturbation of the atmospheric air circulation system so that the lower troposphere and the Global surface are affected, specifically by $\Delta$T $\sim$ 0.6$^{\circ}$C at 20 km and $\sim$0.2$^{\circ}$C at ground level.

Mention should also be made of the work of Mohakumar (1988) who examined the middle atmosphere (65-70 km altitude) temperature associated with the 11-year solar cycle.  This worker argued that enhanced solar emission of Lyman Alpha (121.6 nm) plays a major role in the physico-chemical processes involving minor constituents.  Lyman Alpha is absorbed mainly at the heights mentioned.
The 10.7 cm radio emission,which is generally regarded as a proxy of the UV flux, and for which there is data back to 1947 (UKSSDC, 2009), is also very relevant to this height region.  We find that, for Cycles 19, 20 and 21, the Odd/Even Cycle maximum of $<$19, 21$>$ to that of 20 in the Solar Radio Flux is 1.8 compared with 1.54 for the peaks of the SSNos.  As remarked already, the hard UV has a bigger peak to peak variation than the sunspots so that if the `1.8' factor itself increases as wavelength increases then this will help with the 11-year, 22-year problem.  It also helps with the positive feedback suggestion for the `high' $\Delta$T-values in Figure 3b.

That this is probably not the whole story, however, comes from the linearity in the plot of $\Delta$T (obs) vs $\Delta$T (pred) - predicted on the basis of SSN, referred to in 2.3.4 (and having a correlation probability $p = 0.000$).  One would have needed a concavity in the plot, ie low SSNs (even numbered cycles) to yield lower UV fluxes (per SSN) than high for the 22-y, 11-y contrast to be due entirely to UV.  If seems likely that there is a big component from the inertia effect already referred to.

At this stage, the dependence of `efficiency of change in temperature for change in SI' as a function of relevant time interval can be extended to include annual and daily variations.  Summarising, we have the following $\Delta$SI/SI percentages needed for a change in temperature of $0.1^{\circ}$C:

1 day  $\sim$  0.6\%, 1 year  $\sim$  0.6\%, together with the values discussed already 11-year  $\sim$  (0.10 $\pm$ 0.02)\% and 22-year  $\sim$  (0.014 $\pm$ 0.004)\%.

The trend seems physically sound.

\subsection{The extent to which the solar irradiance change accounts for Global Warming}

By confirming the conversion factor for changes in SI to changes in temperature we have confirmed the conclusions of our earlier work (Erlykin et al, 2009a) that less than 14\% of the temperature increase (of $0.5^{\circ}$C) between 1956 and 2001 is due to the change in solar irradiance.  Applying the same conversion factor for $\Delta$SI/SI to $\Delta$T from the second half of the Century to the first half, we find that the observed temperature increase (of $0.25^{\circ}$C) can be compared with out prediction of $0.3 \pm 0.1^{\circ}$C.  There is thus no evidence for any excess warming over and above `natural causes' in this period.

\section{Cloud cover, solar irradiance correlations}
\subsection{Post-1984 results : the 11-year cycle}

There is a wealth of literature on the relationship between the mean cloud cover, as deduced from satellite observations (ISCCP) and the SSN or the closely related cosmic ray (CR) intensity.  Much of it relates to `low' clouds (LCC : heights less than 3.2 km), eg Marsh and Svensmark (2000), where, over the 11-y cycle, the peak-to-peak range of LCC is $\sim$2 \%.  For higher clouds, the medium clouds (MCC : 3.2 - 6.5 km) and high clouds (HCC : $>6.5$ km) there are the maps, already referred to, of Voiculescu et al (2006).  An analysis has also be given by ourselves (Erlykin et al, 2009c) in which the CC, SSN correlations have been examined as a function of latitude.  The two analyses are consistent in the sense that there are variations with latitude and the sign of the correlation depends on height.

Specifically, the LCC, SSN (or UV) correlation is negative for LCC (this is why the apparent LCC, CR correlation is positive - since CR and SSN are inversely correlated).  The MCC, SSN correlation is positive, and the HCC, SSN correlation has equal (small) areas of positive and negative correlation (Voiculescu et al, 2006).  However there is a bigger area having a negative HCC, CR correlation and this may have arisen from a mis-identification of SSN and CR, (actually UV and CR)), in which case the HCC, SSN correlation is also positive like that for MCC, SSN.

\subsection{Longer period correlations : post-1952}

As remarked earlier, synoptic cloud data are available for the last half-century and these are very useful for the present analysis. They comprise measurements from a very large number of sites distributed over the oceans.  Some 60 million observations were involved in the period 1954 - 2000.  Every effort was taken to ensure that the same criteria were adopted by the observers.  Unfortunately, the observations are divided only into `high clouds' and `low clouds' and thus a comparison with the HCC, MCC and LCC is difficult; however, it can and will be done.

Figure 5 shows the results for the CC magnitudes, as a function of time, for the two latitude regions, both over the oceans.  The SSN data are averaged over 11-year cycles as in Erlykin et al (2009a), but the CC are not(an unimportant fact), but rather to means over a 5 year bin.  The random error on each point is about $\pm$ 0.15\%.  There is seen to be a strong anti-correlation of CC with SSN for the latitude range 30$^{\circ}$ - 60$^{\circ}$~N and a lesser one for 30$^{\circ}$ S - 30$^{\circ}$ N, both for low clouds. These results are in the same sense as for our own work on LCC, where, as with Marsh and Svensmark (2000), we found a positive correlation of LCC with CR - in the form of a negative correlation of LCC with SSN (CR and SSN being inversely correlated).  This follows because we found that the positive LCC, CR averaged over all latitudes was greater than the negative MCC, CR correlation averaged in the same way.  Thus LCC plus a fraction of MCC will have a positive CC, CR correlation. There is also a marked upward trend in the CC with time.

For high clouds, the correlation is in the same sense as we found for the 11-year analysis of HCC and MCC.  Here, for high clouds there is a modest downward trend in CC with time.  The opposite trend with time for `low' and `high' clouds is in the spirit of our suggestion referred to earlier of the inverse correlation in general of LCC and MCC.

It is interesting that the `1970 dip' so strongly marked in the $\Delta$T plot (Figure 3b), is associated with the peaks in low cloud and the minima in high cloud, both in Figure 5. This again agrees with the fact that there is a strong negative correlation between LCC and MCC; indeed our explanation of the apparent LCC, CR correlation itself is that it is SI (rather than CR) that is responsible, the mechanism being the heating of the earth's surface causing a change in mean cloud height (by about 40m from 1985 - 2005), which caused the LCC to fall and the MCC to rise.

Concerning the 11-year cycle in the synoptic cloud data, although none is visible in the low, high and latitude-divided data in Figure 5 it is readily apparent in the work of Pall\'{e} Bag\'{o} and Butler (2000), who combined the data of Norris (1999).  The authors found a peak-to-peak variation for a three-year running mean of the yearly daytime total cloud cover over the ocean of $0.7 \pm 0.2\%$.

A slow rise in low cloud cover over the 50-year period with a somewhat smaller fall in high cloud cover over the same period is a prominent feature of Figure 5.  We estimate an overall mean increase of $3.0 \pm 0.5\%$ over both cloud height ranges and both latitude ranges.  Pall\'{e} Bag\'{o} and Butler's value is $2.5 \pm 0.5\%$.  Some confirmation comes from the summary of cloud data by Bryant (1997) who finds increases of $\sim 2\%$ over each of the N and S oceans and $3\%$ over Europe.  Interestingly, it is found that there is a peak in cloud cover in 1945 for the N and S oceans, Europe and Canada, with a peak in 1950 for the USA (but none for Australia and India).

Presumably they are connected with `our' temperature peaks in 1940 and 1950 (see Section 2.3.2).

It can be added that Pall\'{e} Bag\'{o} and Butler (2000) find support for the cloud cover trend from sunshine records over 4 sites in Ireland, and provide arguments why such records should have wider application.

\subsection{Discussion of the Cloud Data}

Concerning the 11-year cycle there is consistency in the sense that the sign of the correlation of CC with SSN changes phase with increasing altitude for both the maps of Voiculescu et al (2006) and our own analysis (Erlykin et al, 2009b).  The reason for the `sign' reversal Haigh (2007), for temperature, at least, in which the Solar UV causes changes in the atmospheric air circulation system and thereby mean air temperature versus height.  In this model the magnitude of the 11-year peak-to-peak temperature cycle increases with height systematically, unlike in the cloud case, where the CC `amplitude' falls with height, as well as oscillating in sign.  The answer might be the role of ice crystals, which have different contributions as a function of height.  The same situation may be responsible for the difference between the observed and expected latitude variation of CC.

Turning to the `22-y oscillation', the observation of the same pattern in both the 30$^{\circ}$ - 60$^{\circ}$ N region and that for 30$^{\circ}$ S - 30$^{\circ}$ N, albeit with reduced magnitude, is reassuring. Also reassuring is the accuracy with which the minimum SSN in 1970  and maximum in 1985 is reproduced for the High Cloud (30$^{\circ}$ - 60$^{\circ}$ N) and the inverses for the Low Cloud (30$^{\circ}$ - 60$^{\circ}$ N).  Explanation of the cloud pattern in terms of systematic errors is surely ruled out. Turbulent effects associated with the equatorial region can be invoked to explain the reduced peak-to-peak variation, in the near Equatorial regions.

The long-term drift visible in both Figures 5a and 5b can be considered next.  Norris (2000) has argued that the trend may be spurious but, in our view, the consistency between the two latitude ranges and the analyses by other authors, argues against this likelihood.  Comparison of the trends can be made with the ISCCP data.  The latter show, for LCC, a distinct fall of mean cloud cover with time from 1984 onwards and this is opposite to the trend found for the synoptic cloud results for `low' clouds.  It would be surprising if the difference in the `low' definitions had such a big effect.  More relevant is to examine the trend for the whole cloud amount, ie `low' and `high' (synoptic) with LCC + MCC + HCC.  The former is still positive (Figures 5a and 5b) at the rate of about +3\% per 30 years.  For the ISCPP data the overall trend is -3\% per 20 years.  There is a clear discrepancy.  The situation of CC with respect to temperature will be discussed next.

\section{The relationship of surface temperature and cloud cover}

It is generally accepted that low clouds are negatively correlated with surface temperature because of their degree of absorption of incoming radiation and high clouds have a positive correlation because of their `greenhouse effect', ie reflection of terrestrial infra-red radiation.

Comparison of Figures 3b and 5a, b show that for the 22 y `cycle' this is indeed the case; the temperature dip in 1970 corresponds to the high value of low cloud cover in Figure 5a and the low value of high cloud cover in Figure 5b.  This is a satisfying result.

Less satisfying is the situation with the long term trends in CC in Figures 5a and 5b.  It will be remembered that the upward temperature trend has been removed from Figure 3b.  Thus, we have an overall temperature rise associated with a rising low cloud cover and a slightly falling high cloud cover, ie opposite to the situation for the 22-year modulation.  Although by no means certain, it is likely that this situation arises because the `Global Warming' is due to anthropogenic materials which do not result in changes to the cloud cover in the same sense as that for SI changes.  The manner in which the magnitude of the CC change depends on the `time interval' can be considered, in a similar way to that for temperature (Section 2.4.2).

Summarising, we have the following $\Delta$SI/SI changes needed to change the CC by $2\%$.

\begin{center} 
\begin{tabular} {l l l}
1 day & 29\% (night/day, typically) & 29\%\\
1 year & (latitude variation of ISCCP data) & 6\%\\
11 year & (LCC) & 0.1\%\\
22 year & (50y analysis, here) & 0.01\%\\
\end{tabular}
\end{center}
As with the temperature changes, the trend seems reasonable.

\section{The last 20 years}

The SSN has been anomalously low for several years and currently shows no sign of increasing!  The associated CR intensity may have at last reached its maximum, at least there is a flattening in the contemporary neutron monitor rates at some locations, McMurdo and Inuvik, sites in the Arctic region where particle of the lowest energy - and which are most affected by the solar wind - are able to reach the Earth along the magnetic field lines (Bartol neutron monitors, 2009).  However other sites show a continuing rise.  There is agreement, however, that the CR intensity is higher than has ever been previously recorded (eg Calgary neutron monitor, 2009).

The CR peak is interesting in that normally the SSN would have started to rise up to a year previously (the well known hysterysis effect caused by CR diffusion) but it did not.  A further anomaly has been the unusually low peak SSNo for Cycle 23 - the peaks for odd numbered cycles have usually been higher than the mean of their neighbours, as can be seen in Figure 2.  Interestingly, the UV profile for Cycle 23 was normal.  The CR time-profile for the Cycle was also anomalous (Ahluwalia and Ygbuhay; 2008) in the sense that there was a `shoulder' in the neutron monitor count rate in 2004 and 2005.

All these facts make us aware of a possible anomaly in the temperature record (Figure 6) - there was a dramatic drop in $\Delta$T for 2008.  Specifically, $\Delta$T fell by 0.20$^{\circ}$C whereas an increase of $\sim$ 0.1$^{\circ}$C might have been expected, since anthropogenic heating is supposed to be gaining ground.  It might be thought conceivable that all the anomalies were related and that the 0.2$^{\circ}$C cooling was indicative of a mechanism related to the prolonged very low SSN, or the forecast by Ermakov et al (2009) involving the frequency distribution of past temperature periodicities and the involvement of periodicities for the arrival of cosmic dust - which should have implications for climate.

It is relevant at this stage to point out that there have been forecasts of `Global Cooling'.  For example, Landscheidt (2000) found a correlation of Global temperature with the strength of the solar wind and used the fall in wind strength (as indicated by the `aa' geomagnetic parameter) after 1990 to predict a cooling, after a phase - lag of some 4-6 years.  Clearly, such a cooling did not start as predicted (in 1994-1996) but an open mind should be kept about the cooling prospect.

The statistical significance of the 2008 reduction is not high, however, for the following reasons:

\begin{itemize}
\item[(i)] The dispersion of yearly values about the smoothed temperature variation 
with time (shown dashed) is as indicated `obs' in Figure 6.If many factors contribute 
to the temperature variation in a random way (~a rather uncertain assumption, although 
there are, in fact, several independent contributors~) the probability of the 2008 
deviation can be estimated. It is at the 2 standard deviations level, ie a 2.5\% 
probability.  
\item[(ii)] A 0.2$^{\circ}$C or more dip has occurred on 16 occasions in the period of the observations, 128 y, ie a probability per year, if random, of 12.5\%
\item[(iii)] In fact, there are corrections to be applied to the temperature record for ENSO, volcanoes, ozone and greenhouse gas fluctuations, as described earlier.  The standard deviations for the 5-y and 1-y corrections (none of which have been applied to the data) are also shown in Figure 6.  If added to the (obs.) fluctuations in quadrature, the significance of the 2008 dip falls to about one standard deviation, ie $\sim$ 16\%, assuming a Gaussian distribution.
\item[(iv)] The `noise' correction (ENSO + etc) is not, in fact, completely random, largely because of the near-periodicity in ENSO.  This feature is presumably responsible for the fact that most of the minima in Figure 6 are separated by 2y.  In a sense, a reduction in 2008 might have been predicted from known mechanisms.
\end{itemize}

Taking all the factors above into account, no case can be made for the 2008 minimum having an anomalous origin.

It can be remarked that the shallow convexity in the dashed curve is consistent with the `standard' SI effect.

\section{Conclusions}

The relation between changes in solar irradiance and changes in mean surface temperature and cloud cover at various levels has been examined.  The comparatively high temperature changes associated with changes in the solar irradiance (as evinced by change in sunspot numbers) are confirmed; changes in UV as distinct from longer wavelengths are a strong candidate. Positive feedback may also be a contributory process, such as occurs in the Arctic where an increased surface temperature melts ice and reduces the albedo so that the temperature rise is enhanced. A similar situation pertains in the stratosphere. The increase in $\Delta$T with respect to expectation as a function of `integration time' - 1y, 11y and 22y - points to an inertial effect, ie the time-constant of the atmosphere/Earth's surface temperature system, but it must be said that the difference between the 22-year and 11-year temperature variations - a factor $\sim 5$ - is rather dramatic.  Our result is bigger than the factor 1.7 found by Miyahara et al (2008) (for 26-year and 12-year cycles) from tree ring data over the last 500 years.

It might be thought that the 22-year cycle temperature change was due to a solar wind effect, in view of the well-known 22-year cycle of the solar magnetic field direction.  Indeed, a cosmic ray origin might even be postulated.  However, inspection of the CR induced atmospheric ionization versus time (eg Bazilevskaya et al, 2008) does not show an adequate 22-year modulation.  Instead, we prefer an intrinsic-to-the sun origin, involving ultra-violet radiation together with positive feedback.

The changes in cloud cover correlate appropriately with the temperature changes on the 20-30 y scale but the mechanism is still unclear.  The longer term slow increase in low cloud cover (and small reduction in high cloud cover) is of the opposite `sign' to expectation.  An explanation in terms of anthropogenic causes for the temperature rise seems likely, although it cannot be ruled out that the slow cloud cover changes are an artifact.

The lack of an explanation for the actual geographical pattern of the strong correlation of low cloud cover with solar irradiance (negative)(see Section 1), is still present.  An explanation in terms of the changes to the atmospheric circulation `geography' not being as predicted by the model of Haigh (2007) is a distinct possibility, although the magnitude of the effect may well be as has been estimated.

The mechanism responsible for the temperature and Cloud Cover changes is clearly `solar' but whether the initiating energy is supplied by radiation (UV, as described) or whether it is the solar wind is not yet clear.  However, in view of the energy in the solar wind being only of order one millionth of that in sunlight, the solar wind hypothesis has difficulties: a `positive feedback' of the magnitude required would appear to be very unlikely.

Our estimate of `less than 14\% for the period 1956-2002 is confirmed.  For the previous 50 years, changes in solar irradiance appear to be responsible.

Thus, extra forcing (presumably anthropogenic) started to become important only in the 1950s.  This conclusion confirms that of Lean et al (2005), and others (notably that of the IPCC).

\section{Acknowledgement}

We express gratitude to the Dr John C Taylor Charitable Foundation for financial support of this project. Dr Joan Kenworthy is thanked for helpful comments on the manuscript.

\newpage

{\bf Table 1. Sources of the data for Figure 1 : temperature change, $\Delta$T, versus change in solar irradiance, $\Delta$SI}\\

\begin{tabular}{| l | p{5cm} | p{5cm} | l |}\\ \hline
\textbf{`Number'} & \textbf{Remarks} & \textbf{References} & \textbf{Time period}\\ \hline
1 & `The sensitivity of the Earth's surface temperature to changes in irradiance' & Lean et al (2005) from IPCC & $\approx$10 y\\ \hline
2 & The 11-year temperature anomaly & Lean et al (2005); Haigh (2007) & 11 y\\ \hline
3 & The long scale 1970 dip & Erlykin et al (2009a) & 20 - 30 y\\ \hline
4 & Maunder Minimum (variable over the Earth) & Lean et al (2005), mean of range for $\Delta$SI & $\sim$100y\\ \hline
5 & Decadel variations of sea surface temperatures (very variable) & White et al (1997) Van Loon et al (2007) & 10 y\\ \hline
6 & Orbital changes (calculations for 65$^{\circ}$N, for periods with no magnetic field changes) & Rusov et al (2008) & 10$^{5}$ y\\ \hline
7 & Global seasonal - summer, winter temperature differences vs latitude & $\Delta$T data from Allen (1973) & 1 y\\ \hline
8 & Changes over the last 10$^{4}$ y & IPCC (1990) & 10$^{4}$ y\\ \hline
9 & Change since the last Ice Age & Lean et al (2005) & 2.10$^{4}$ y\\ \hline
10 & The average from Figure 3b & The present work & 20-30 y\\ \hline
\end{tabular}

\newpage

{\bf Table 2. Correlation probabilities and slopes of the line, $\Delta$T (obs.) vs $\Delta$T (predicted) for the period 1000-1900AD. The `correlation probability' is the chance of a random correlation giving the observed value or greater; the number of standard deviations from zero for the slope is another indicator of the significance of the correlation.}\\

\begin{tabular} {|l | l | l |}\\ \hline
\textbf{Period} & \textbf {Correl. prob.} & \textbf {Slope}\\  \hline
1000 - 1100 & 0.327 & 0.36 $\pm$ 0.20 \\ \hline
1100 - 1200 & 0.167 & 0.63 $\pm$ 0.28 \\ \hline
1200 - 1300 & 0.000 & 1.55 $\pm$ 0.15 \\ \hline
1300 - 1400 & 0.037 & 0.399 $\pm$ 0.092 \\ \hline
1400 - 1500 & 0.013 & 1.06 $\pm$ 0.21 \\ \hline
1500 - 1600 & 0.013 & 0.883 $\pm$ 0.25 \\ \hline
1600 - 1700 & 0.248 & 0.379 $\pm$ 0.37 \\ \hline
1700 - 1800 & 0.021 & 0.922 $\pm$ 0.19 \\ \hline
1800 - 1900 & 0.07 & 1.11 $\pm$ 0.18 \\ \hline
\end{tabular}

\newpage

\section{References}
\begin{itemize}

\item[1.] Ahluwalia,H.S. and Ygbuhay,R.C., 2008. Observed Galactic Cosmic Ray 11-year modulation for Cycle 23, 30th Int. Cosmic Ray Conf., Merida, Mexico, 493.
\item[2.] Allen,C.W., 1973. Astrophysical Quantities, Athlone Press, Univ. of London.
\item[3.] Bartol Neutron Monitors: (2009) $<$http://neutron.bartol.udel.edu.html$>$
\item[4.] Bazilevskaya,G.A. et al (12 authors), 2008. Ionization Processes in Planetary Atmospheres Cosmic Ray Induced Ion Production in the Atmosphere, Space Science Reviews 137, 1.
\item[5.] Bryant,E., 1997. Climate processes and Change, Cambridge University Press.
\item[6.] Calgary Neutron Monitor: (2009)\\ 
$<$ftp://ftp.ngdc.noaa.gov/STP/SOLAR-DATA/COSMIC-RAYS/calgary.tab$>$
\item[7.] Crowley,T.J., 2000. Causes of Climate Change over the past 1000 years, Science, 289, 270.
\item[8.] Deland,M.T. and Cebula,R.P., 2008. Creation of a composite solar ultraviolet irradiance data set, submitted to Journal of Geophysical Research available from $<$http://lasp.colorado.edu/lisird/delandcomposite/index.html$>$
\item[9.] Eichler,E., Olivier,S., Henderson,K., Laube,A., Beer,L., Papina,T., Gaggeler,H.W., and Schwikowski,M., 2009. Temperature response in the Altai region lags forcing, Geophysical Research Letters, 36, L01808.
\item[10.] Erlykin,A.D., Sloan,T. and Wolfendale,A.W., 2009a. Solar Activity and the Mean Global Temperature, Environmental Research Letters, 4, 014006.
\item[11.] Erlykin,A.D., Gyalai,G., Kudela,K, Sloan,T. and Wolfendale,A.W., 2009b. `On the correlation between cosmic ray intensity and cloud cover', Journal of Atmospheric and Solar-Terrestrial Physics, DOI: 10.1016/j.jastp.2009.06.012; arXiv:0906.4442
\item[12.] Erlykin,A.D., Sloan,T. and Wolfendale,A.W., 2009c. `Correlations of clouds, cosmic rays and solar irradiation over the earth', Solar Physics (submitted for publication)
\item[13.] Foukal,P., North,G. and  Wigley,T., 2004. A stellar view on solar variations and climate. Science 306, 68.
\item[14.] Foukal,P., Frohlich,C., Spruit,H., and Wigley,T.M.L., 2006. Variations in solar luminosity and their effect on the Earth's climate. Nature, 443, 161.
\item[15.] Gray,L.J.,  Rumbold, S.T. and Shine, K.P., 2009. Stratospheric temperature and radiative forcing response to 11-year solar cycle change in irradiance and ozone. DOI:10.1175/2009JAS 2866.1.
\item[16.] Haigh,J.D., 2007. The Sun and the Earth's Climate, Living Rev. Solar Physics 4, 2.
\item[17.] Hansen,J.E., Ruedy,R., Sato,M., Imhoff,M., Lawrence,W., Easterling,D., Peterson,T., and  Karl,T., 2001. A closer look at United States and global surface temperature change, Journal of Geophysical Research 106, 23947, updated by GISS, NASA 2009-01-09.
\item[18.] Hegerl,G.C. et al, 2007. `Understanding and Attributing Climate Change 2007', CUP, UK and NY.
\item[19.] Hood,L.L., 1997. The solar cycle variation of total ozone : dynamical forcing in the lower stratosphere. Journal of Geophysical Research 102, 1355.
\item[20.] Hood,L.L. and Soukharev,B.E., 2000. The solar component of long-term stratospheric variability: observations, model comparisons and possible mechanisms.  Proc. SPARC 2nd General Assembly November 6-10.
 \item[21.] IPCC: Climate Change, The IPCC Scientific Assessment, Ed Houghton,J.D., Jenkins,G.J. and Ephraums,J.J., Cambridge University Press, (1990).
 \item[22.] IPCC : Climate Change 2007: The Physical Basis. Contribution of Working Group 1 to the Third Assessment Report of the International Panel on Climate Change, Cambridge University Press, Cambridge, UK and New York, USA.
\item[23.] The ISCCP data were obtained from the International Cloud Climatology Project website : http://isccp.giss.nasa.gov/ maintained by the ISCCP research group at NASA Goddard Institute for Space Studies, New York, Rossow,W.B. and Schiffer,R.A., 1999. `Advances in understanding clouds from the ISCCP', Bulletin of the American Meteorological Society 80, 2261.
\item[24.] Khristjansson,J.E. and  Kristiansen,J., 2000. Is there a cosmic ray signal in recent variations in global cloudiness and cloud radiative forcing?, Journal of Geophysical Research 105(11), 851.
\item[25.] Landscheidt,T., 2000. The Solar Cycle and Terrestrial Climate, ESA SP-463.
\item[26.] Lean,J., Rottman,G., Harder,J. and  Kopp,G., 2005. Source contributions to new understanding of global change and solar variability, Solar Physics 230, 27.
\item[27.] Marsh,N. and Svensmark,H., 2000. Low cloud properties influenced by cosmic rays, Physical Review Letters. 85, 5004.
\item[28.] Meehl,G.A., Washington,W.M., Ammann,C.A., Arblaster,J.M., Wigleym,T.L.M. and  Tebalid,C., 2004. Combinations of Natural and Anthropogenic Forcings in the Twentieth-Century Climate, Journal of Climate 17, 3721.
\item[29.] Miyahara,H., Yokoyama,Y. and  Masuda,K., 2008. Possible link between multi-decadal climate cycles and periodic reversals of solar magnetic polarity.  Earth and Planetary Science Letters 272, 290.
\item[30.] Mohakumar,M., 1988. Response to an 11-year solar cycle on Middle Atmosphere Temperature, Physica Scripta 37, 460.
\item[31.] Nagashima, T., et al, 2006. The effect of carbonaceous aerosols on surface temperature in the mid twentieth century, Geophysical Research Letters 33, L04702.
\item[32.] Norris,J.R., 2004. Low Cloud Type over the Oceans from surface observations, Part II : Geographical and Seasonal Variations, Journal of Climate II (1998), 383, and "Change in Near-Global Cloud Cover and Reconstructed Radiation Flux since 1952". Private communication.
\item[33.] Norris,J.R., 2000. What can cloud observations tell us about climate variability, Space Science Reviews 94, 375.
\item[34.] Pall\'{e} Bag\'{o},E. and Butler,C.J., 2000. Sunshine, clouds and cosmic 
rays. star.arm.ac.uk/rambn/345.pdf.
\item[35.] Peixoto,J.P. and Oort,A.H., 1992. Physics of Climate, American Institute of 
Physics.
\item[36.] Rusov,V. et al, 2008. Galactic Cosmic Rays - Clouds effect and bifurcation model of the Earth's global climate. Private Communication (siiis@te.net.ua)
\item[37.] Shindell,D., Rind,D., Balachandran,N., Lean,J. and Lonergan,P., 1999. Solar cycle variability, ozone and climate, Science 284, 3050-58.
\item[38.] Sloan, T. and Wolfendale,A.W., 2008. Testing the proposed causal link between cosmic rays and cloud cover, Environmental Research Letters 3, 024001.
\item[39.] UKSSDC, 2009 : ftp://ftp.ukssdc.ac.uk/wdc/a.w.wolfendale@durham.ac.uk/.
\item[40.] Vecchio,G.L., and  Nanni,T., 1995. The atmospheric temperature in Italy during the last 1000 years and its relationship with solar output, Theoretical and Applied Climatology 51, 159.
\item[41.] Van Loon, H., and Shea,D.J., 2007. A probable signal of the 11-year solar cycle in the troposphere of the northern hemisphere, Geophysical Research Letters 26, 2893.\item[42.] Viereck,R.A. and Puga,L.C., 1999. The NOAA MgII core-to-wing index: Construction of a 20 year time series of chromospheric variability from multiple satellites, Journal of the Geophysical Research 104, 995-10006.
\item[43.] Voiculescu,M., Usoskin,I.G. and Mursula,K., 2006. Different response of clouds to solar input, Geophysical Research Letters 33, L21802.
\item[44.] Wang,Y-M, Lean,J.L., Sheeley Jr.,N.R., 2005. `Modeling the Sun's Magnetic Field and Irradiance since 1913', Astrophysical Journal 625, 522-538.
\end{itemize}
\newpage

\section{Captions to Figures}

{\bf Figure 1}
Changes in mean surface temperature of the Earth ($\Delta$T) versus changes in the solar irradiance ($\Delta$SI).  The sources of the data are given in Table 1. `Model' is expectation from the observation that for all planets the mean surface temperature varies as (SI)$^\frac{1}{4}.$

{\bf Figure 2}
Sunspot number versus time since 1990.  The downward arrows represent periods when the long term mean was low and where (in Figure 3b) we identify temperature dips. The Cycle numbers are indicated along the top.

{\bf Figure 3a}
Average surface temperatures versus time for the whole Globe (GLO), land (LAN) and the oceans (OCE).  The shaded areas represent predictions for specific models, the upper regions include anthropogenic changes and the lower regions without.  The Figures are from Hegerl et al (2007).

{\bf Figure 3b}
Change in temperature ($\Delta$T) and sunspot number (SSN) smoothed (over 11-years) from Erlykin et al (2009a); the smooth slow increase in temperature, ie the `Global Warming', has been removed. We use the SSN as the proxy for SI. $\Delta T_{corr}$ is the change in temperature after correction by us for the effects of volcanoes, ENSO, ozone and greenhouse gases.

{\bf Figure 4} $\Delta$T vs $\Delta$SSN for the 5 dips shown in Figure 3b, with 3 pairs of values for each dip.  The number (10, 30 ...90) is the year (1910, 1930...) for the minimum temperature in that Figure. The dashed line shows the best fit.  The probability of obtaining a `zero-fit' is 0.009, confirming a good correlation.

{\bf Figure 5a}
Changes of cloud cover ($\Delta$CC) for low clouds for the latitude ranges indicated, and associated changes in sunspot number ($\Delta$SSN).  The cloud data are from Norris (2004).

{\bf Figure 5b}
Changes of cloud cover ($\Delta$CC) for high clouds for the latitude ranges indicated, and associated changes in sunspot number ($\Delta$SSN).  The cloud data are from Norris (2004).

{\bf Figure 6}
$\Delta$T versus year for the last 20 years (from Hansen et al, 2001, updated by GISS, NASA, 2009)    The dashed curve is the 5-year average. `obs' denotes the median spread of the yearly points about the mean line. (5) and (1) refer to the median spread of the corrections (which have not been applied) for ENSO, volcanic forcing, etc.

\newpage

\begin{figure}[htbp]
\begin{center}
\epsfxsize=15cm\epsfbox{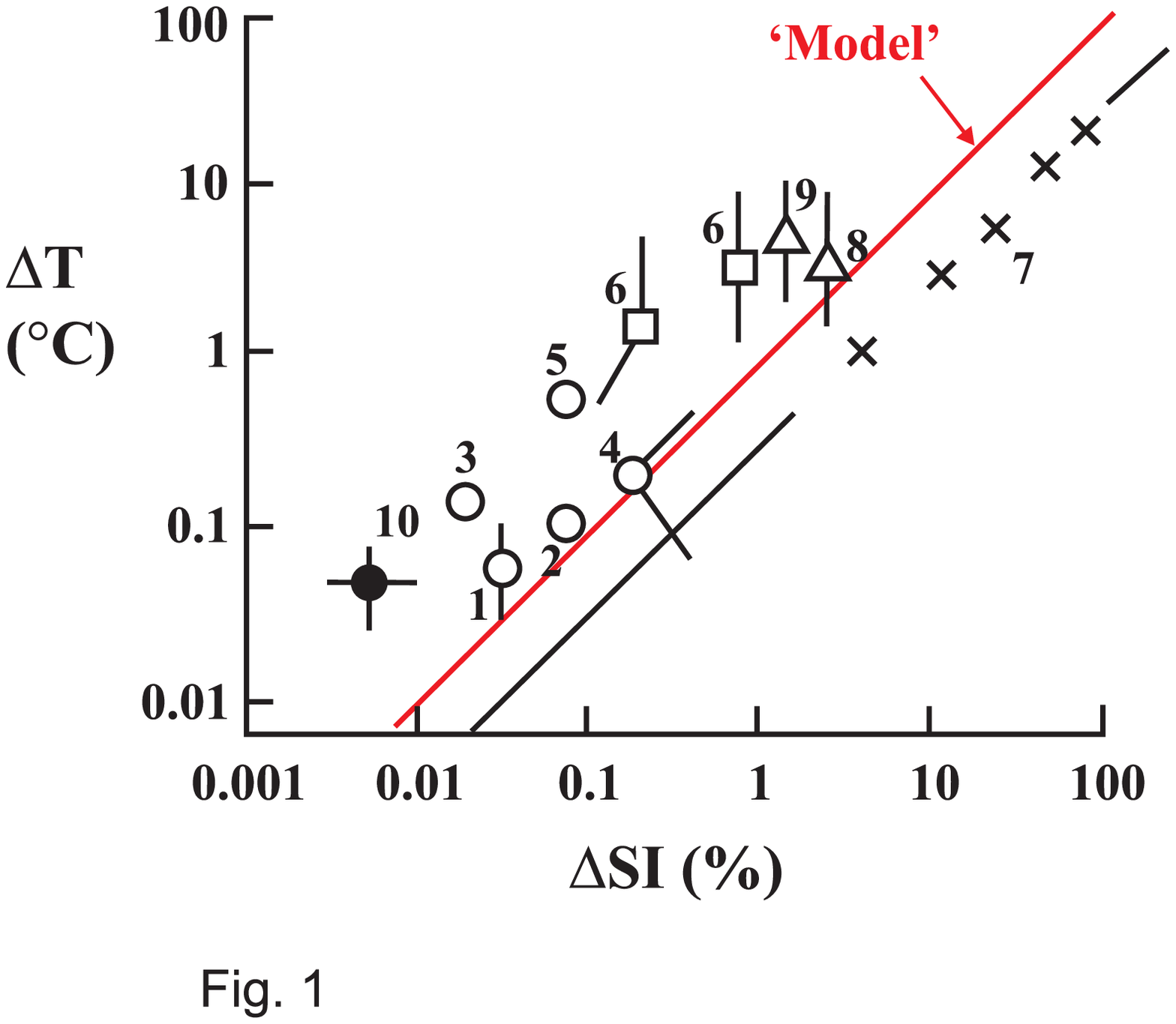}
\end{center}
\end{figure}

\newpage

\begin{figure}[htbp]
\begin{center}
\epsfxsize=15cm\epsfbox{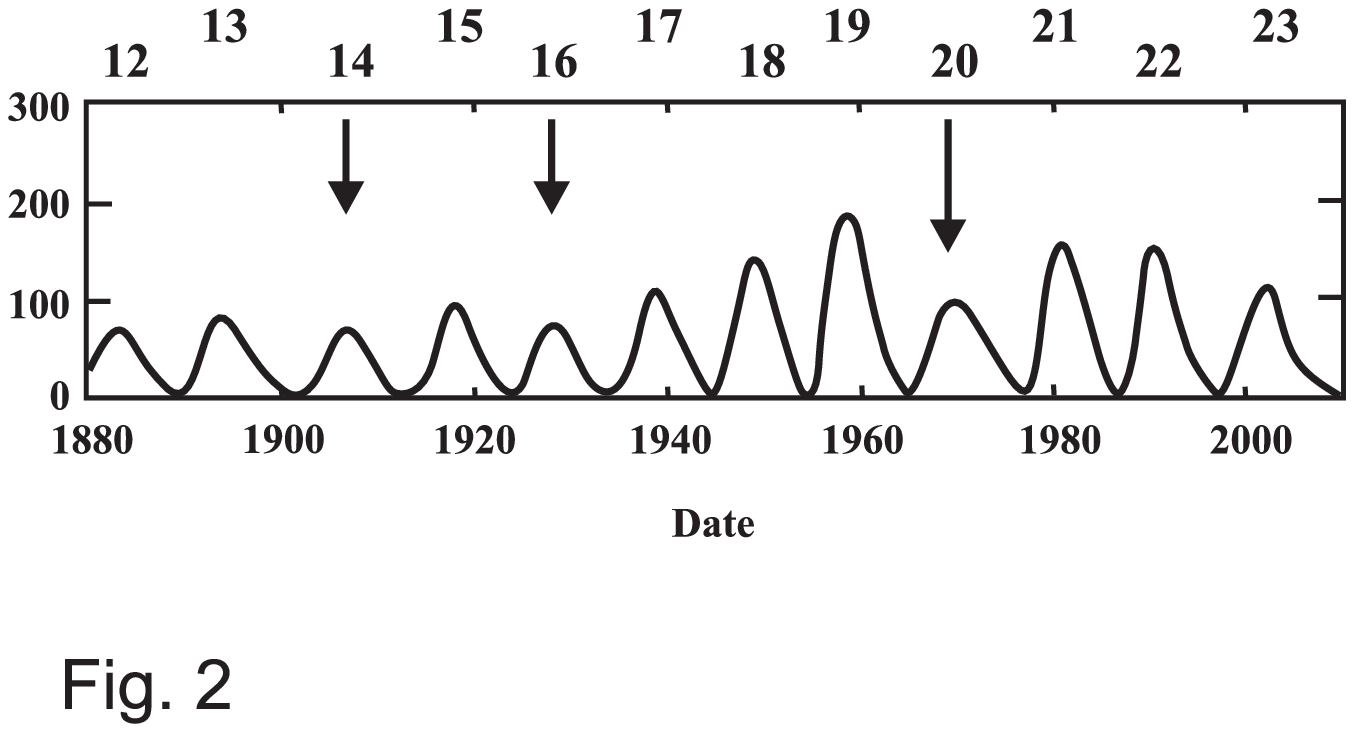}
\end{center}
\end{figure}

\newpage

\begin{figure}[htbp]
\begin{center}
\epsfxsize=15cm\epsfbox{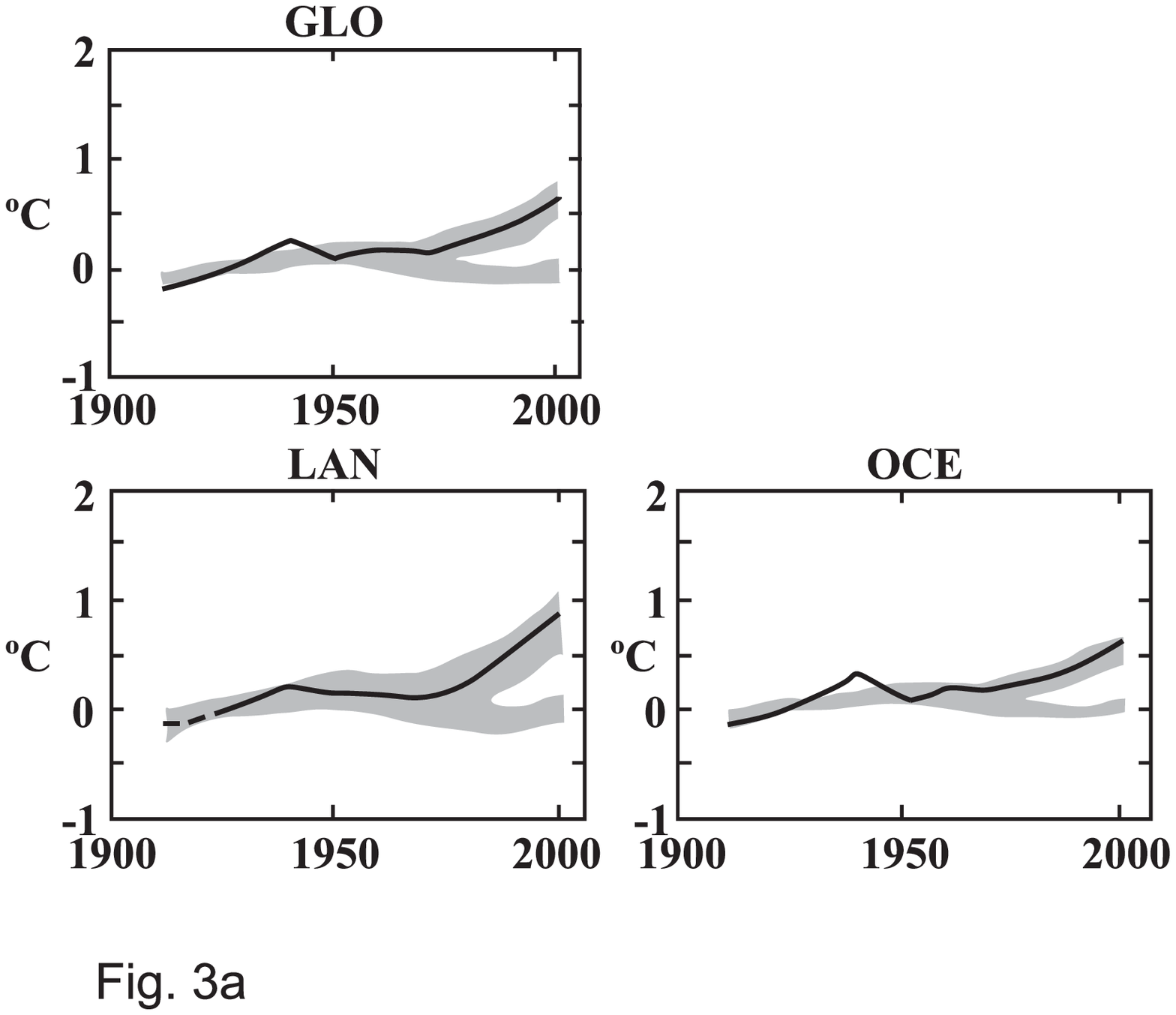}
\end{center}
\end{figure}

\newpage

\begin{figure}[htbp]
\begin{center}
\epsfxsize=10cm\epsfbox{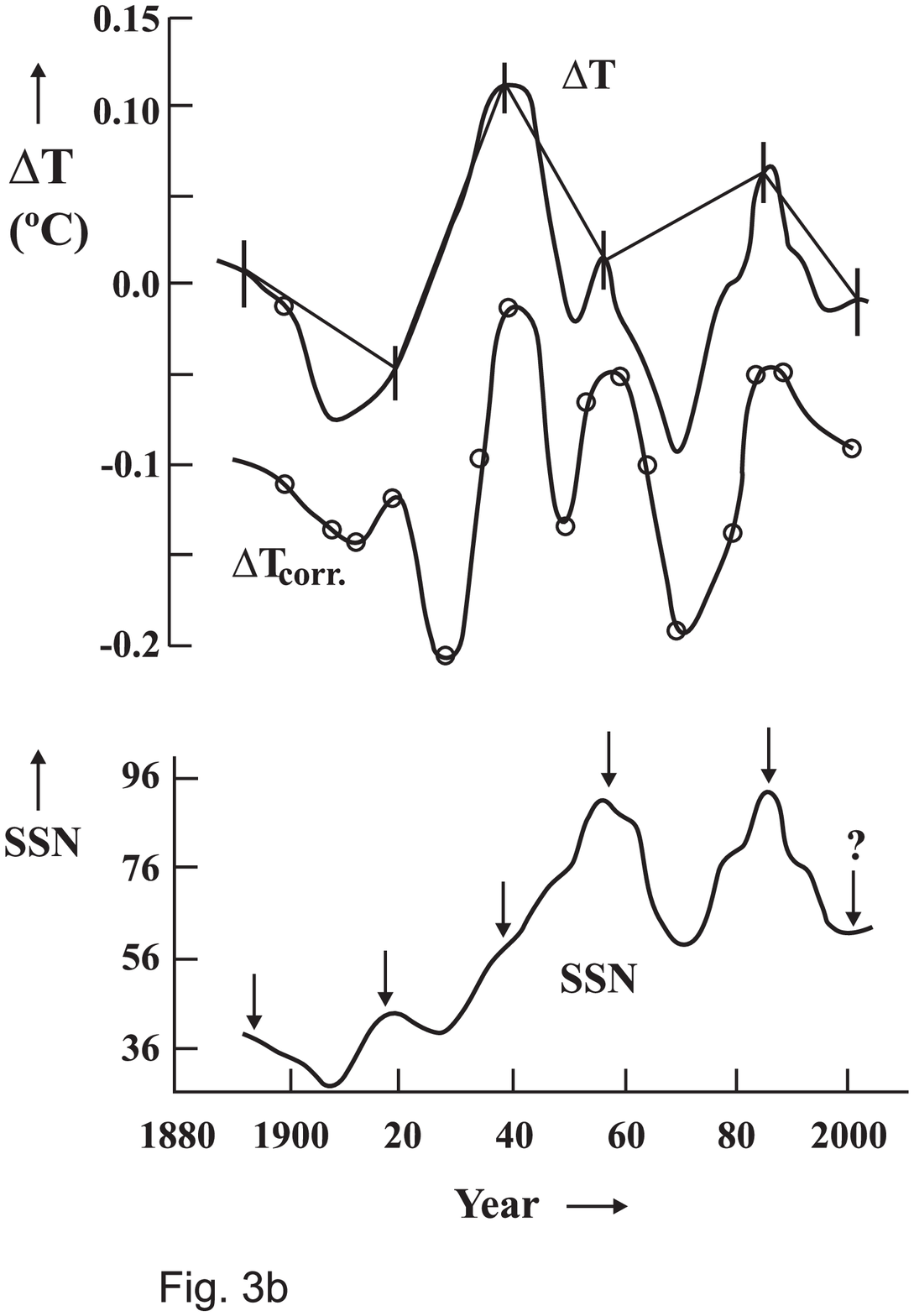}
\end{center}
\end{figure}

\newpage

\begin{figure}
\begin{center}
\epsfxsize=10cm\epsfbox{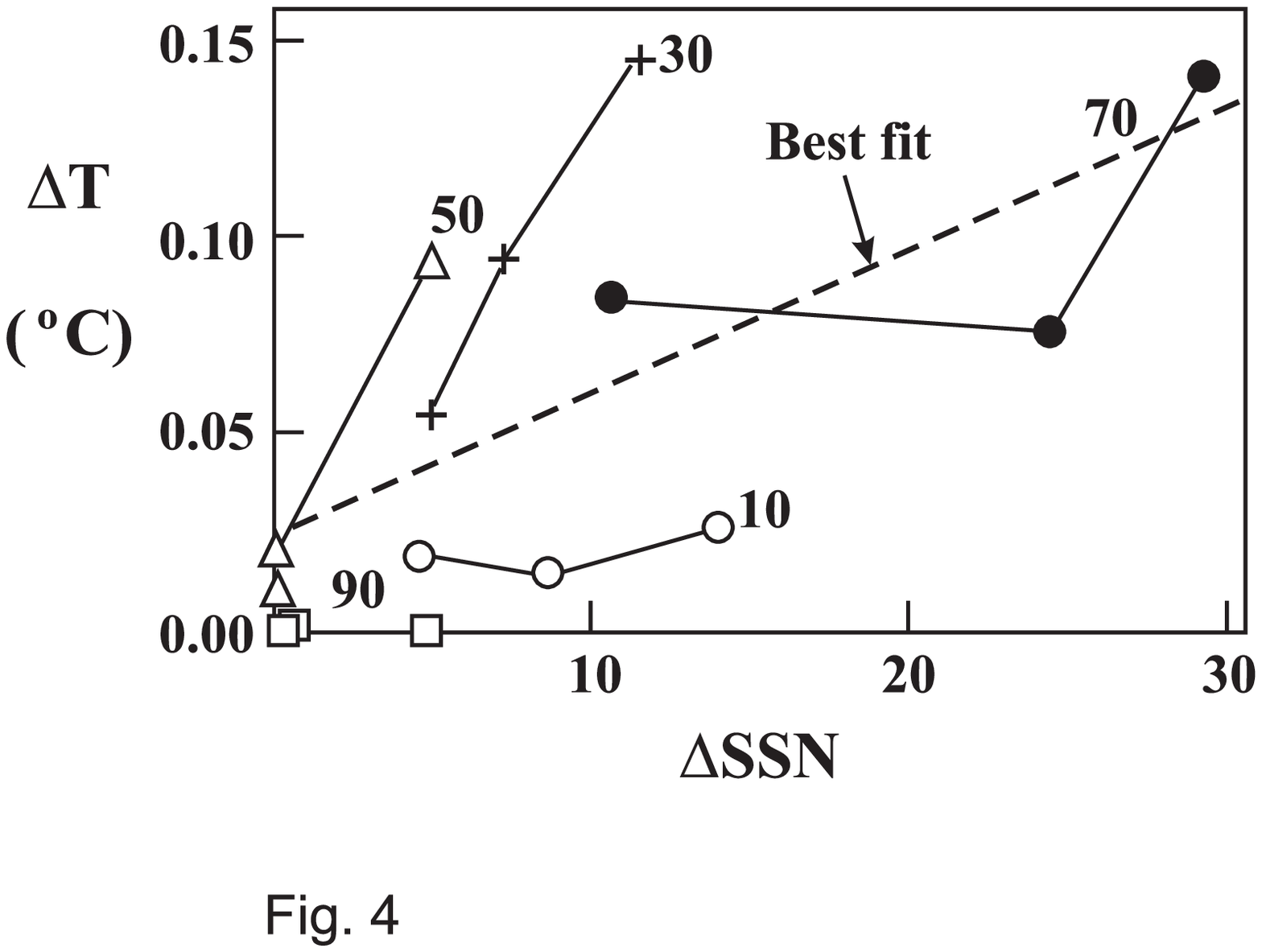}
\end{center}
\end{figure}

\newpage

\begin{figure}
\begin{center}
\epsfxsize=15cm\epsfbox{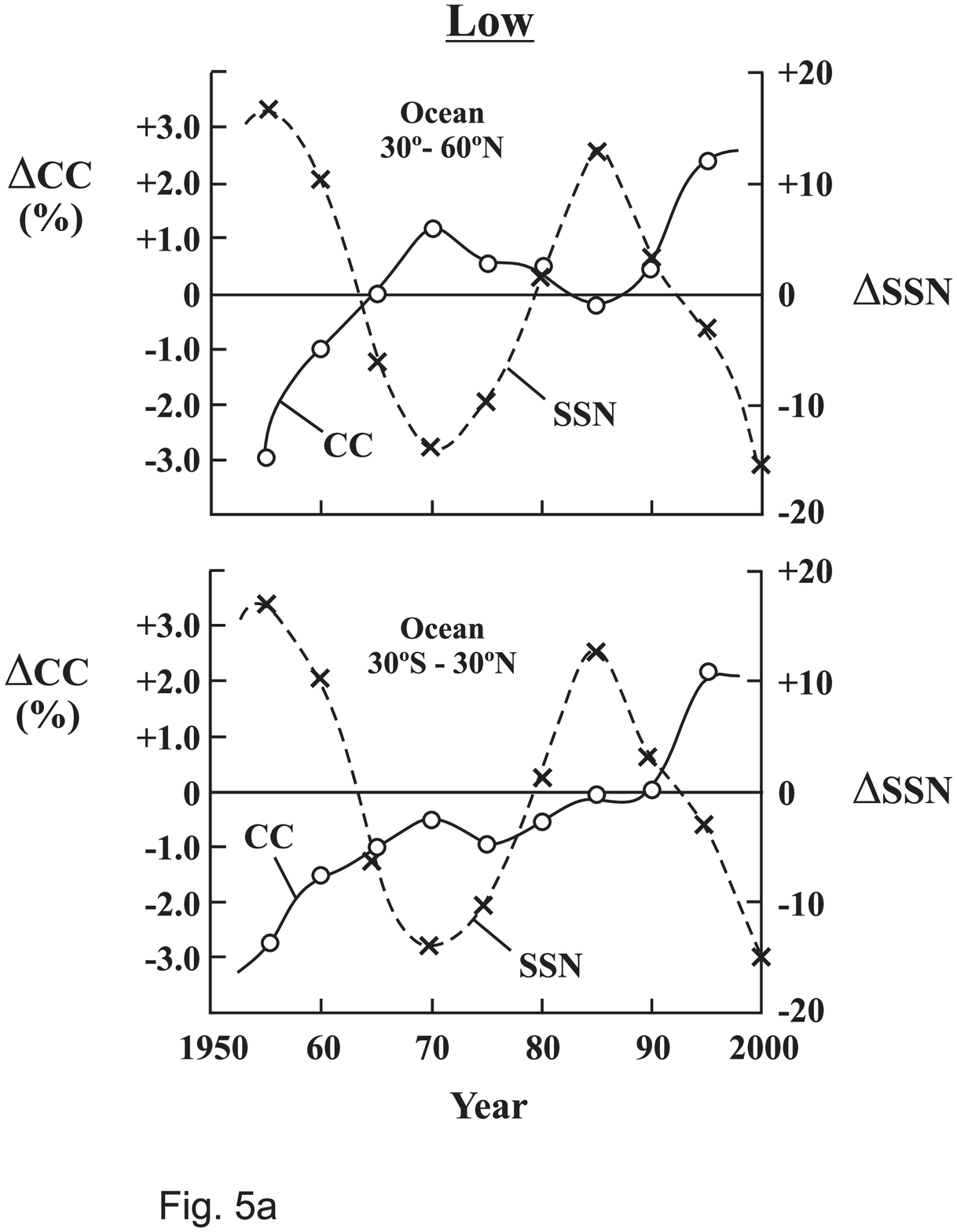}
\end{center}
\end{figure}

\newpage

\begin{figure}
\begin{center}
\epsfxsize=15cm\epsfbox{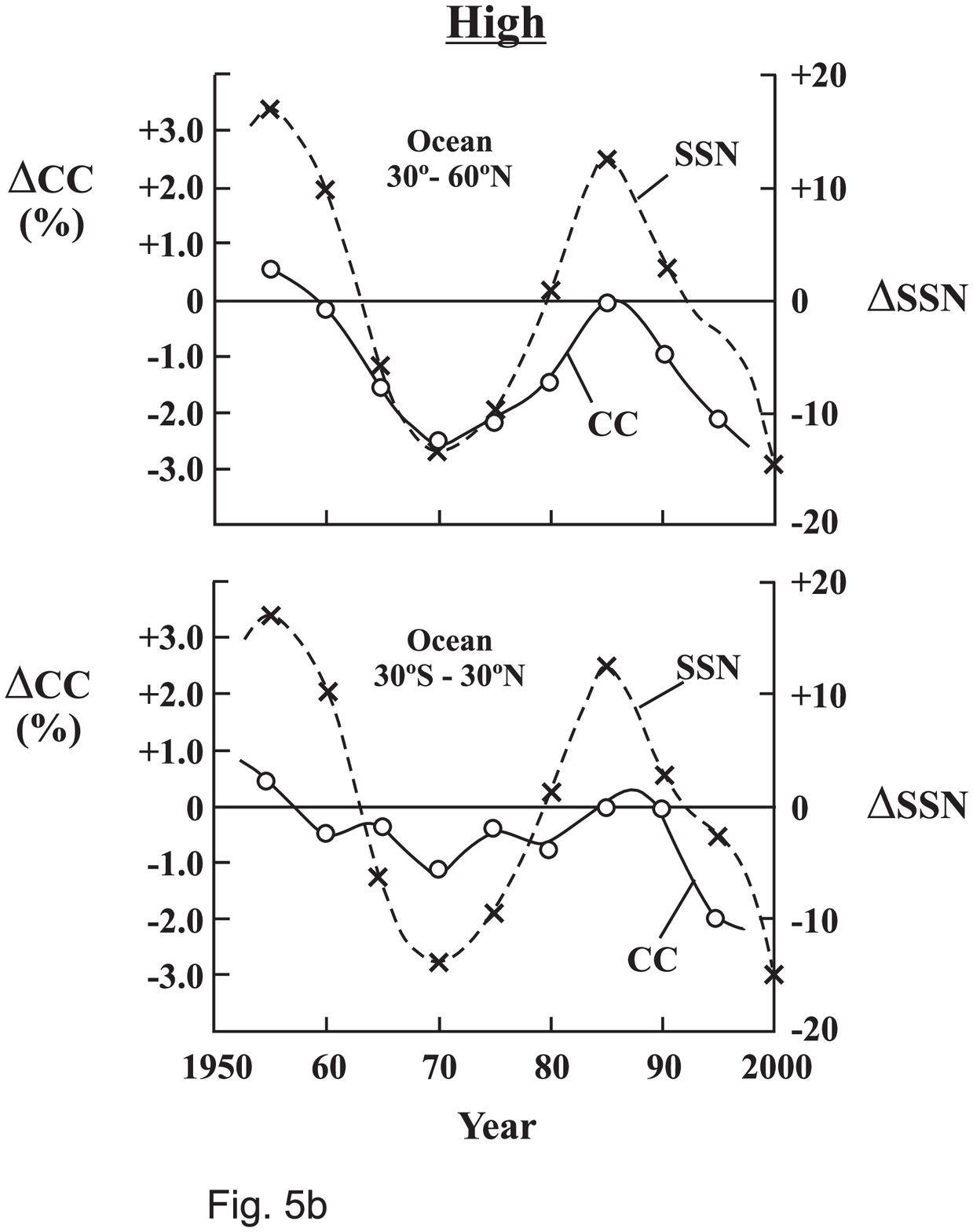}
\end{center}
\end{figure}

\newpage

\begin{figure}
\begin{center}
\epsfxsize=15cm\epsfbox{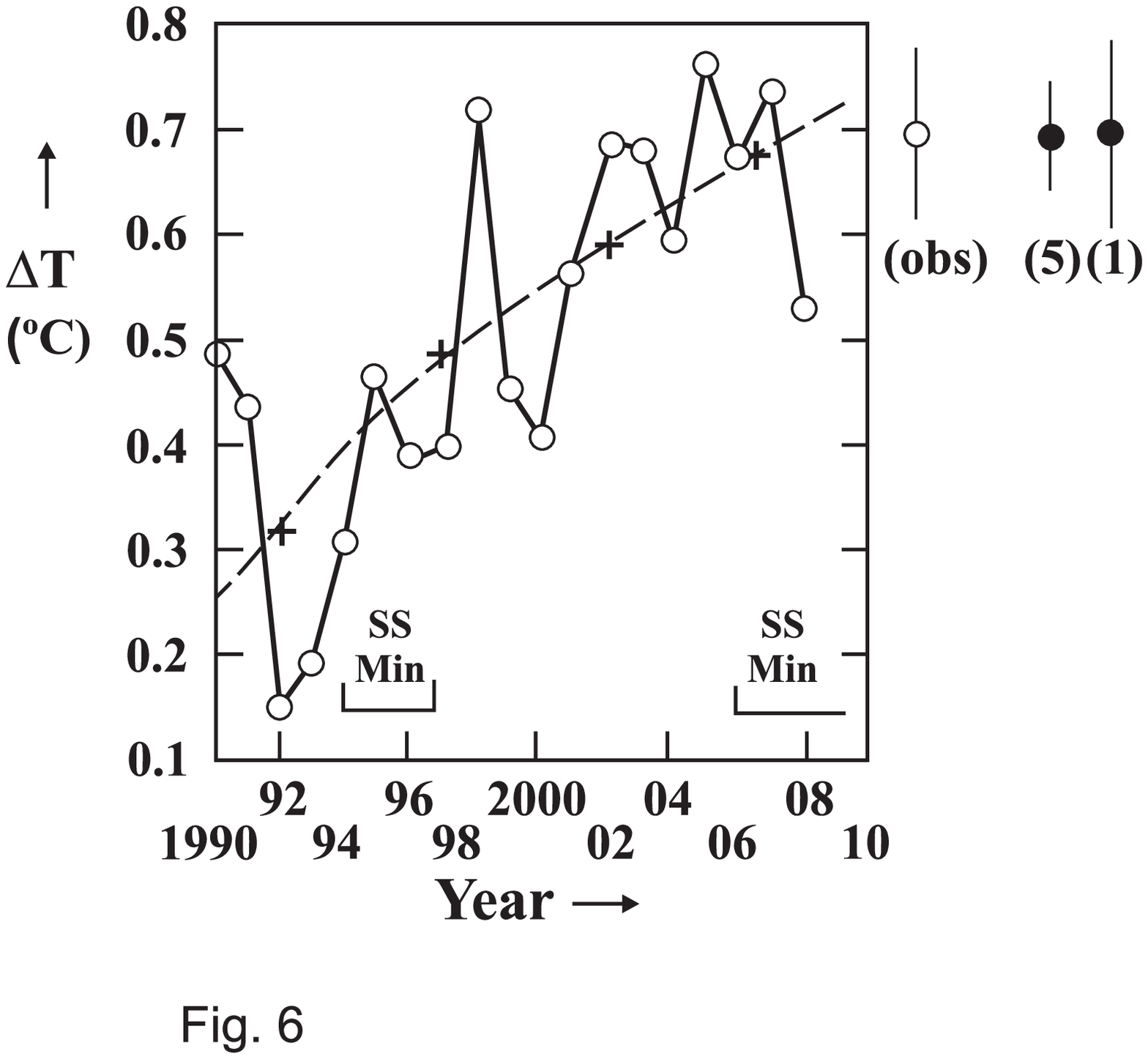}
\end{center}
\end{figure}

\end{document}